\documentclass[amsmath,prl,aps,twocolumn]{revtex4}
\usepackage{amsthm,amsfonts,graphicx,verbatim}

\newcommand{\be}{\begin{equation}}
\newcommand{\ee}{\end{equation}}
\newcommand{\bea}{\begin{eqnarray}}
\newcommand{\eea}{\end{eqnarray}}

\renewcommand{\vec}[1]{{\bf #1}}
\renewcommand{\epsilon}{\varepsilon}

\begin{document}

\title{Dispersion interaction between crossed conducting wires}
\author{John F. Dobson${}^{1}$, Timothy Gould${}^{1}$ and Israel Klich${}^2$}
\affiliation{${}^1$ 
Nanoscale Science and Technology Centre, Griffith University,  Nathan Queensland 4111, Australia, and \\
CSIRO National Hydrogen Materials Alliance, CSIRO Energy Centre, \\
10 Murray Dwyer Circuit, Steel River Estate, Mayfield West, NSW 2301, Australia \\
 ${}^2$
Department of Physics, University of Virginia, Charlottesville, VA
22904 }



\begin{abstract}
We compute the $T=0K$ Van der Waals (nonretarded Casimir) interaction energy $E$ between two infinitely
long, crossed conducting wires separated by a minimum distance $D$ much greater than their radius. We find that, up to
a logarithmic correction factor, $E\propto -D^{-1}\left| \sin
\theta \right| ^{-1}f(\theta )$ where $f(\theta )$ is a
smooth bounded function of the angle $\theta$ between the wires. We recover a conventional result of the form
$E\propto -D^{-4}\left|\sin\theta \right| ^{-1}g(\theta )$ when  we include an electronic energy
gap in our calculation.  \ Our prediction
of gap-dependent energetics may be observable
experimentally for carbon nanotubes, either via AFM detection 
of the vdW force or torque, or indirectly via observation of mechanical oscillations. 
This shows that strictly parallel wires, as assumed in previous predictions, are not 
needed to see a novel effect of this type.
\end{abstract}
\maketitle


At the micro- and nano-scale, dispersion (van der Waals, vdW) forces are ubiquitous
\cite{Isra},
and recent advances in manufacturing and measurement techniques have prompted much interest in their precise form.

The simplest theories sum vdW
interaction energies between pairs of molecules, which is a
good approximation for dilute insulating objects, where the dipole
fluctuations at different points of one body are almost independent. For
non-dilute dielectric and magnetic materials this summation
approximation can be misleading, and may even give the wrong
sign of the interaction \cite{Kenneth02}. Moreover, 
for anisotropic conducting nanostructures, the non-locality of 
Coulomb screening and associated density correlations within each 
object may change the form of the dispersive forces altogether \cite{Dobson06,CylsPlates}.

This physics is exemplified by the class of quasi-1D objects, 
which exhibit correlation phenomena of both theoretical and
experimental interest. Indeed, in the extreme limit
truly confined $1D$ electrons experience 
a Luttinger liquid instability, as 
(e.g.) in single walled
armchair nanotubes \cite{Egger}.

When two such "wires" are placed parallel and close to each other, the coulomb
interaction between their density fluctuations may become a relevant perturbation,
resulting in rich behavior at low
temperatures and densities, such as locked charge density waves and 
Wigner cristallization. \cite{Klesse00}. The density density interaction is also
responsible for coulomb drag phenomena whereby a current applied to
one wire induces voltage on the other wire \cite{Hu,Nazarov}.

Dispersion forces between 1D systens are interesting 
even at larger separations. For the case of two infinitely long strictly 
parallel wires separated by distance $D$ greatly exceeding their
 radius $b$, the interaction energy is known \cite{Dobson06,Chang71,TanAnderson,Matloob,Drummond07}
to be strongly dependent on the presence of an electronic energy gap: 

\begin{eqnarray}
E^{vdW}/L\propto -D^{-2}(\ln (D/b))^{-3/2} (metallic).
\label{EParallelConductors} \\
E^{vdW}/L\propto -D^{-5}  (semiconducting) \label{EParallelInsulators}
\end{eqnarray}%

The result (\ref{EParallelConductors}) was obtained via zero-point
energies of coupled plasmons in the random phase approximation (RPA),
followed by a perturbative evaluation of the resulting integral
 \cite{Chang71,Dobson06}. 
The form (\ref{EParallelConductors}) was also supported by diffusion Monte Carlo calculations \cite{Drummond07}.

Such forces may be important when one
considers solutions of nanotubes or long molecules. Indeed, while
"wires" minimize their energy by aligning, in a solution of
nanotubes a parallel configuration might not be formed because of
entropic reasons. Therefore, it is important to understand the
wire-wire interaction for a general orientation.  
Here we consider the angle and distance dependence for wires that are well
separated.  We will show that the (absolute) vdW energy of a pair of
 non-parallel wires inclined at angle $\theta$ is, up to a logarithmic 
correction factor specified later,  

\begin{eqnarray}
E^{vdW}\propto -D^{-1} \left| \sin \theta \right| ^{-1}
f(\theta) \;\;\;  (metallic) \label{QuoteResultMet} \\
E^{vdW} \propto -D^{-4} \left| \sin \theta \right| ^{-1}g(\theta) \;(semiconducting)
\label{QuoteResultsemicon}
\end{eqnarray}
Here $D$ is the least distance between points on the two wires,
and $\theta $ is the angle between the wires.  $f$ and $g$ are smooth bounded functions.
 In this limit there is also a prospect of measuring the 
dispersion force between two nanotubes directly, via Atomic Force Microscopy (AFM) or spectroscopy of mechanical vibrations. Eqs (\ref{QuoteResultMet},\ref{QuoteResultsemicon}) show that unusual
gap-dependent results are expected from such experiments not only when the tubes are parallel \cite{Dobson06},
but also when they are non-parallel.   

The crossed-wire interaction has previously been related to the interaction between
anisotropic media in a similar manner to the way Casimir forces
between molecules can be obtained from the Lifshitz formula by
taking the dilute limit. The interaction between nonisotropic
materials related to our problem was studied in \cite{Kenneth98}:
there, the energy of a pair of media conducting only in
prescribed (but different) directions is computed. In
\cite{Rajter07} the interaction between non-isotropic dielectric
media was considered and the limit of a dilute medium was
associated with the interaction between a pairs of wires. 
The asymptotic results of \cite{Rajter07} were similar to those of 
a pair summation \cite{Parsegian} approach (i.e. like Eq. (\ref{QuoteResultsemicon})) 
even for metallic cases), and qualitatively different from Eq. (\ref{QuoteResultMet}))
 derived below for metallic wires.
We suspect that this difference may be due to the
different Coulomb screening physics of a 
single pair of metallic wires compared to an infinite array of such wires. (See
 \cite{Gould09} for similar considerations relating to layered systems).
This question is related to the very non-additive dispersion 
physics of low-dimensional, zero-gap systems in general \cite{Dobson06}. 

In \cite{Emig08}, the orientational
interaction between spheroidal dielectrics has been computed,
 but the metallic limit treated here was not yet considered.
Some related work has also
been done on 1D conductors in a collinear ''pointing''
configuration \cite{White08}, but the calculation appropriate for
metallic 1D conductors does not appear to have been done for cases
where they are aligned in a non-parallel geometry.

Here we will compute the dispersion energy
directly for crossed wires within second order perturbative non-retarded theory, as justified below. 
The interaction energy in this approach is (see e.g. \cite{Zaremba76}) 

\begin{eqnarray}
E^{(2)}&=-\frac{\hbar }{2\pi }\int_{0}^{\infty }du\int
d^{3}r_{1}d^{3}r_{2}d^{3}r_{1}^{\prime }d^{3}r_{2}^{\prime
}V(r_{12})V(r_{1^{\prime }2^{\prime }}) \nonumber\\
  &\times \chi^{A}(\vec{r}_{1},\vec{r}_{1}^{\prime },iu)\chi ^{B}(\vec{r}_{2},\vec{r}_{2}^{\prime },iu)
\label{Zaremba76} \\
&=-\hbar (2\pi )^{-7}\int d\vec{p}d\vec{p}^{\prime }\bar{V}(\left| \vec{p}%
\right| )V(\left| \vec{p}^{\prime }\right| )\nonumber \\
&\times \int_{0}^{\infty }du\chi ^{(A)}(-%
\vec{p},-\vec{p}^{\prime },iu) 
\chi ^{(B)}(\vec{p},\vec{p}^{\prime },iu)
\label{ZKInFourierSp}
\end{eqnarray}
Here $\chi(\vec{r},\vec{r}')$ is the density-density response of each
 sub-system in the absence of the other sub-system, and 
$\chi(\vec{p},\vec{p}^\prime)$ is its 3D Fourier transform.  
$V(\vec{r})$ is the  bare Coulomb interaction between electrons in the two subsystems, 
and $V(\vec{p})=4\pi e^2/|p|^2$ is its 3D Fourier transform. 
V can be considered small in the present case because of the large
 inter-wire separation assumed here. Indeed it is easily verified that 
(\ref{Zaremba76}) yields (\ref{EParallelConductors}) for the case of parallel
well-separated 1D conductors.

\begin{figure}
\includegraphics*[scale=0.4]{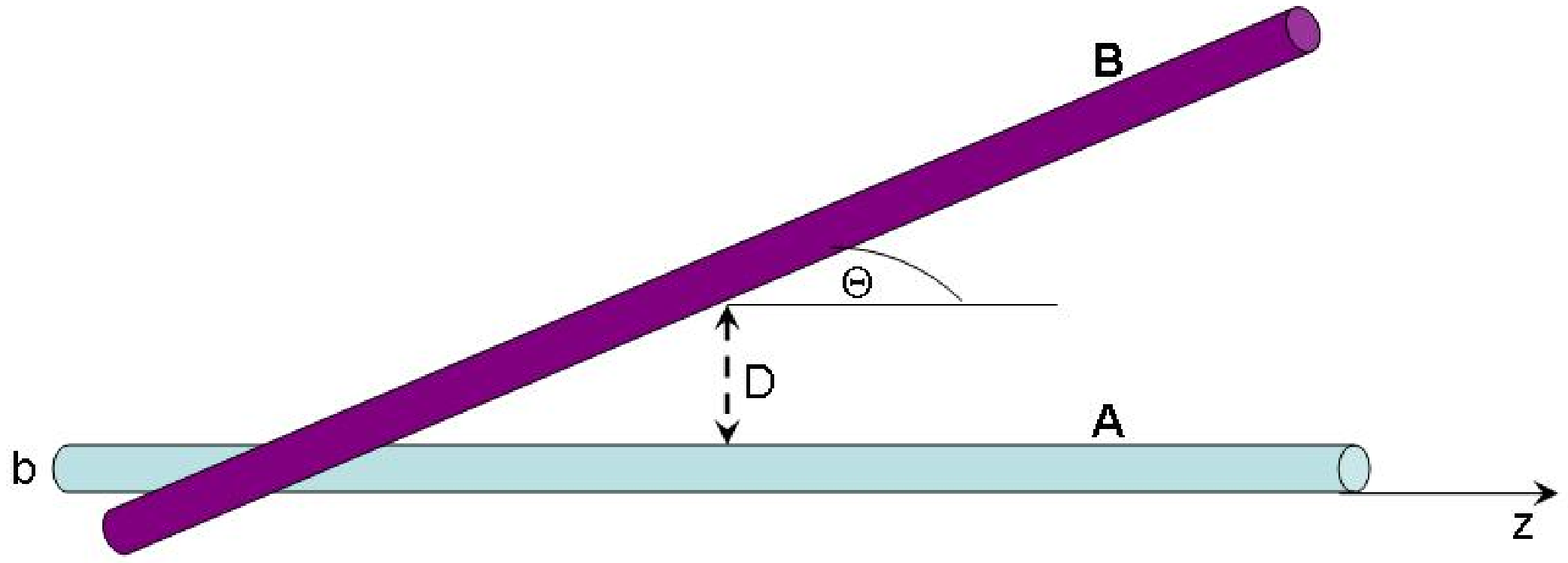}
\caption{Van der Waals interaction of two crossed wires $A$ and
$B$ at distance $D$ and angle $\theta$} \label{fig1}
\end{figure}

We consider a pair of wires $A,B$ described in Fig.\ref{fig1}. One
wire is assumed to lie along the $z$ axis. The xz plane and the
origin are determined by demanding that the point on wire B, lying
closest to wire A, is the point $(D,0,0).$ The orientation of wire
B is then determined by a single angle $\theta .\;$

For later convenience we introduce unit vectors in the yz plane,
parallel and
perpendicular to wire $B$: \;%
$\hat{u} =(0,\sin \theta ,\cos \theta )$,  
$\hat{v} =(0,\cos \theta ,-\sin \theta )$.
Locations along wire A are denoted $z$, and locations along wire B
are determined by a signed 1D position variable $s$ so that 
general points on wires A and B are \;%
$\vec{R}_A(z)=(0,0,z)$,\; \; $\vec{R}_B(s)=D\hat{x}+s\hat{u}$ 

The density-density response functions in 3D space of the two wires are written in
terms of an assumed strictly 1D density-density response function
 $\bar{\chi}(z,\omega)$ for electron
motion along a wire:

\begin{eqnarray}
\chi ^{(A)}(\vec{r},\vec{r}^{\prime },\omega
)&=\bar{\chi}(z-z^{\prime },\omega )\delta (x)\delta (x^{\prime
})\delta (y)\delta (y^{\prime }) \label{ChiA}\\
\chi ^{(B)}(\vec{r},\vec{r}^{\prime },\omega )
&=\bar{\chi}(\left( \vec{r}-\vec{r}^{\prime }\right)\cdot\hat{u},\omega)
 \nonumber\\
&\times\delta (x-D)\delta (x^{\prime }-D)\delta (\vec{r}\cdot \hat{v})\delta (\vec{r}%
^{\prime }\cdot \hat{v})
\label{ChiB}
\end{eqnarray}
We express $\bar{\chi}(z,\omega)$
in terms of its 1D Fourier transform.
$\bar{\chi}(z,\omega )=\frac{1}{2\pi }\int \bar{\chi}(q,\omega)\exp
(iqz)dz
$.
Then, doubly Fourier-transforming (\ref{ChiA}) and (\ref{ChiB}) we obtain

\begin{eqnarray}
\chi ^{(A)}(\vec{q},\vec{q}^{\prime })&=2\pi \int
dk_{1}\bar{\chi}(k_{1},\omega )\delta (q_{z}+k_{1})\delta
(q_{z}^{\prime }-k_{1})\;\;, \label{ChiAqqp} \\
\chi ^{(B)}(\vec{q},\vec{q}^{\prime })&= 2\pi \int
dk_{2}\bar{\chi}(k_{2},\omega )\exp (i(q_{x}+q_{x}^{\prime })D) \nonumber \\
&\times 
 \delta (\cos \theta q_{z}-\sin \theta
q_{y}+k_{2})\delta (\cos \theta q_{z}^{\prime }-\sin \theta
q_{y}^{\prime }-k_{2}) \nonumber \\
\label{ChiBqqp}
\end{eqnarray}
Putting (\ref{ChiAqqp},\ref{ChiBqqp}) into (\ref{ZKInFourierSp}), and defining
\begin{eqnarray}
F(k_{1},k_2)&=&\int_{0}^{\infty
}\bar{\chi}(k_{1},iu)\bar{\chi}(k_{2},iu)du
\label{DefF} \\
V_{2D}(D,p)&=&2\pi  e^2 \exp(-pD)/p \label{DefV2D}
\end{eqnarray}
we obtain
 \begin{eqnarray}
E^{(2)} &=-\frac{\hbar }{(2\pi )^{5}}\int
dk_{2}dk_{1}d^{3}pd^{3}p^{\prime }F(k_1,k_2)
V(\left| \vec{p}\right| )V(\left| \vec{p}^{\prime }\right| )%
\nonumber \\
&\times \exp (i(p_{x}+p_{x}^{\prime })D)\delta
(k_{1}-p_{z})\delta
(p_{z}^{\prime }+k_{1}) \nonumber \\
&\times  \delta (\cos \theta p_{z}-\sin \theta p_{y}+k_{2})\delta
(\cos \theta p_{z}^{\prime }-\sin \theta p_{y}^{\prime }-k_{2})  \nonumber\\
=-\frac{\hbar }{\left( 2\pi \right) ^{3}}&\int dk_{1}
dk_{2}{F(k_{1},k_{2})\over \sin ^2 \theta}  
V_{2D}\Big(D,\sqrt{k_{1}^{2}+(-\frac{k_{1}\cos \theta +k_{2})}{\sin
\theta })^{2}}\Big)^2  .\nonumber \\ 
\label{EZKAs2Dkintegral}
\end{eqnarray}%
We calculate the single-wire response $\bar{\chi}$ for eqs (\ref{DefF},\ref{EZKAs2Dkintegral})
within the Random Phase Approximation.
From the work of Li and Das Sarma \cite{Li89},\; 
this RPA approach for the long-wavelength collective motions 
can even be applied to a strongly interacting model of a 1D conducting system
such as a Luttinger liquid.  This would not apply for broken-symmetry cases such as 
a Wigner-crystalline wire, but this case is likely to occur only for small inter-wire 
separations \cite{Klesse00}, which are not the focus in the present work.

The starting point of the RPA for a single wire is the $bare$ density-density response $\bar{\chi} _{0}(k,\omega)$. 
The simplest model for $\bar{\chi} _{0}$ 
, at small $k$ and small $\omega =iu$, is%
\begin{equation}
\bar{\chi} _{0}(k,iu)=-\frac{k^{2}n_{01D}}{m(u^{2}+\omega _{0}^{2})},
\label{Bare1DChi0}
\end{equation}%
where $n_{01D}$ is the number of electrons per unit length in the
ground state, and $m\,$is an appropriate mass. $\ \omega _{0}$
(with $\hbar \omega _{0}\approx $ Bloch energy gap $Eg$) is a
harmonic pinning term allowing the modelling of both
semiconducting and metallic wires. \ We take $\omega _{0}=0$ for
the metallic case, and $\omega _{0}\neq 0$ for semiconductors.
(\ref{Bare1DChi0}) follows from classical motion of individual
independent electrons ($F=ma$) but is also valid for the quantal
motion of independent Fermions at small wavenumber
$k$, corresponding, in the present case, to large interwire
separation $D\sim k^{-1}$. 

The Coulomb interaction
$e^{2}/\left| z_{1}-z_{2}\right| $ between electrons on a strictly
one-dimensional wire has a divergent Fourier transform
corresponding to the region $z_{1}\rightarrow z_{2}$. In real
quasi-1D structures such as carbon nanotubes there is a finite
radius $b$ corresponding to the lateral spatial extent of the
one-electron orbitals. \ This serves to smear the Coulomb
divergence at $z_{1}\rightarrow z_{2}$.  The result depends on the 
detailed cross section assumed.  For definiteness we choose a smeared
Coulomb potential for electrons free to move on a cylindrical shell of
zero thickness and radius $b$, representing a nanotube.  The smeared
intra-tube potential for this case is analytic: 

\begin{eqnarray}
v_{1D}(k) &= 2 e^2 I­_0(|kb|)K_0(|kb|) \label{VcoulNanotube} \\
& \approx 2 e^2 |ln(|kb|)|,\;\; for |kb|<<1.
\label{vcoul1Duniversal}
\end{eqnarray} 

where $I_0$, $K_0$ are modified Bessels. The long-wavelength form (\ref{vcoul1Duniversal}) is universal to all wire profiles
but causes spurious collective modes and a divergent energy at larger k, requiring a cutoff to be imposed. 
Although the final result is highly insensitive to this cutoff, we preferred to use
 the specific form (\ref{VcoulNanotube}), thereby avoiding a cutoff altogether.

We now use the RPA \ to write the long-wavelength response of
mutually interacting quasi-1D electrons to an external potential
as
\begin{eqnarray}
\bar{\chi} (k,iu)=\frac{\bar{\chi} _{0}(k,iu)}{1-v_{1D}(k)\bar{\chi}
_{0}(k,iu)}=  \frac{n_{01D}k^{2}}{m(u^{2}+\Omega^2(k))}
 \label{RPAChiOfWire} \\
\Omega^2(k)=\omega^2_{1D}(k)+\omega_0^2 \label{Omega}
\end{eqnarray}%
Here $\omega _{1D}(k)$ is a quasi-acoustic 1D plasmon frequency,
which can be written in terms of a 1D velocity $c_{1D}$ as
\begin{equation}
\omega _{1D}(k)=c_{1D}\left| k\right| \sqrt{I_0(|bk|)K_0(|bk|)
},\;\; c_{1D}=\sqrt{2   e^{2}n_{01D}/m}   \label{Omega1Dfromc1D}
\end{equation}%

Now the 1D plasmon group velocity $\partial \omega_{1D} / \partial q$ from 
(\ref{Omega1Dfromc1D}) is 
much less than the speed of light $c$ except for $q < O(b^{-1}exp(-c^2/c­_{1D}^2))$.
Thus electromagnetic retardation can be ignored except at extremely large distances 
$D>>b \exp (c^2/c_{1D}^2)$.

Puting (\ref{RPAChiOfWire}) into (\ref{DefF})we find
\begin{eqnarray}
F=\left( \frac{n_{01D}}{m}\right) ^{2}\frac{\pi
k_{1}^{2}k_{2}^{2}}{ 2\Omega(k_{1})\Omega
(k_{2})\left\{ \Omega (k_{1})+\Omega
(k_{2})\right\} } \label{ChiChiUIntegral}
\end{eqnarray}

\emph{Case 1, conducting wires, \protect{$\omega_{0}=0 $}}. 

To analyze (\ref{EZKAs2Dkintegral}) we transform to dimensionless plane 
polar coordinates  $(K,\phi)$:
\begin{equation}
k_1 = D^{-1}K sin(\phi +\frac{\theta}{2}),\;\;\;
k_2 = D^{-1}K sin(\phi -\frac{\theta}{2})
\label{PolarCoords}
\end{equation}
with Jacobian $J=D^{-2}K/|sin(\theta)|$.  This yields
\begin{equation}
E^{2:\;metal}=-\frac{\hbar c_{1D}}{16 D |sin(\theta)|} h(\theta , r),\;\;\;r=b/D>>1
\label{EMetFromh}
\end{equation}
\begin{equation}
h(\theta ,r)=\int_{0}^{\infty }dK e^{-2K} \int_{-\pi }^{\pi }d\phi \frac{s_{+} s_{-}}
{\ell _{-}\ell _{+}\left( \ell _{-} s_{-}  + \ell _{+}s_{+}\right) } 
\label{Def_h_theta_r}
\end{equation}
with $s_{\pm} \equiv \left |\sin(\phi \pm \frac{\theta}{2}) \right |$ and
$\ell _{\pm }=\sqrt{I_{0}(Kr s_{\pm})K_{0}(Kr s_{\pm})}$.
Note that $h(\theta ,r)$ is well behaved as $\theta \rightarrow
0$, with a limiting value $h(0,r)=  (ln(D/b))^{-3/2}$ for $r \equiv b/D <<<1$. %

For $r <<<1$ we thus obtain the analytic result
\begin{equation}
E^{(2)}\approx -\frac{1}{16D} \hbar c_{1D}\frac{1}{\left| \sin \theta \right| }\Big(\frac{1}{(\ln
(D/b))^{3/2}}\Big)\;\;\;\;as\;\;\theta \rightarrow 0
\label{SmallTheta}
\end{equation}
For arbitrary $\theta $ and $r \equiv b/D << 1$ the function $h(\theta ,r)$ defined by 
(\ref{Def_h_theta_r}) requires numerical
investigation. We define a smooth function $\Lambda$ with correct angular period:
\begin{eqnarray}
\Lambda(\theta,r)\equiv h(\theta,r)(ln(1/r))^{3/2}\approx \sum_{n=0}^{N}c_n(r)cos(2n\theta) \nonumber \\
c_n(r)\approx \frac{\sum_{i=0}^{\ell}a_{in} x^i}{1+\sum_{j=1}^J b_{jn}x^j},\;\;x=1/lnr 
\label{NumFit}
\end{eqnarray}
The choices $N=5,I=3, J=2$ give a good fit to (\ref{Def_h_theta_r}) for $x<0.4$, i.e. for $D/b \equiv r^{-1} >16.5$, 
and the corresponding coefficients $a,b$ are given in Table 1.

\begin{table}
\begin{tabular}{lllllll}
\protect{$a_{in}$} & n=0 & n=1 & n=2 & n=3 & n=4 & n=5 \\
i=0& 0.85708  & 0.11770 & 0.01534 & 0.00453 & 0.00189 & 0.00097\\
i=1 & 0.18049 & 0.19040 & 0.04743 & 0.01538 & 0.04505 & 0.00228 \\
i=2 & 2.85682 & 0.51752 & 0.12230 & 0.03615 & 0.01378 & 0.00783 \\
i=3 & 2.78324 & 0.61964 & 0.13410 & 0.03497 & 0.03237 & 0.01795 \\
\hline  \\
\protect{$b_{jn}$} & n=0 & n=1 & n=2 & n=3 & n=4 & n=5 \\
j=1 & -2.29966 & -0.24097 & -0.06158 & -0.00820 & -0.00657 & -0.00357 \\
j=2 & 2.47243 & -0.19977 & 0.70113 & 0.91201 & 0.95313 & 0.97461 
\end{tabular}
\caption{Coefficients for log-cosine expansion  \protect{(\ref{NumFit}) of $\Lambda(\theta,r)$}}
\end{table}

\emph{Case 2, semiconducting wires, \protect{$\omega_{0}\ne 0 $}}. 
Here there are two analytic cases according to the separation D: either of the terms on the right of (\ref{Omega}) could dominate for the $k$ values $k\approx D^{-1}$ that dominate the energy integral (\ref{EZKAs2Dkintegral}).

\emph{Case 2a, smaller separations}
When 
\begin{equation}
D<<D_0 \equiv c_{1D}\omega_0^{-1}\sqrt{ln(c_{1D}/(\omega_0 b))} \label{D0}
\end{equation}
we can again ignore $\omega_0$ in (\ref{Omega}), thus recovering the "metallic" vdW energy 
(\ref{EMetFromh}).  Because of our large-separation approximations, this conclusion only holds provided that 
$D_0$ from (\ref{D0}) satisfies $D_0>>b$, which can occur for very small gaps $\omega_0$ such as that noted for (9,3) nanotubes in \cite{Rajter07}.

\emph{Case 2b, larger separations \protect{$D>>D_0$}}
Here we can ignore the $\omega _{P1D}^{2}$ term in (\ref{Omega}) for the relevant values $k\approx1/D$. Then (\ref{DefF}) becomes $
F \approx \left( \frac{n_{01D}%
}{m}\right) ^{2}k_{1}^{2}k_{2}^{2}\frac{\pi /2}{2\omega _{0}^{3}}$
and using (\ref{PolarCoords}), we evaluate (\ref{EZKAs2Dkintegral}) as%
\begin{eqnarray}
E^{(2): semicond} &=&-\frac{3\pi}{1024} \frac{1}{D^{4}}\frac{\hbar
c_{1D}^{4}}{\omega _{0}^{3}}\frac{2\cos ^{2}\theta +1}{\left| \sin
\theta \right| }
\end{eqnarray}
This is consistent with Eq (57) of \cite{Rajter07}.

{\it Conclusions:}  Eq (\ref{EMetFromh}) is the principal result of the present work, along with 
(\ref{Def_h_theta_r}) and 
(\ref{NumFit}).
It shows that nonparallel conducting wires experience a vdW
attractive energy that decays much more slowly with distance
$D$ than the standard $D^{-4}$ dependence predicted by summing
$R_{ij}^{-6}$ contributions over all elements $i,j$ of the wires.
(\ref{EMetFromh}) also shows a strong angular dependence,
giving rise to significant slowly-decaying vdW torques.

The interaction predicted here might be measured directly in
Atomic Force Microscopy experiments on metallic carbon nanotubes,
or indirectly via their mechanical vibrations.
In Figure 2 we estimate the force $F^{met}$ between the conduction-band electrons of two freestanding metallic (5,5) carbon nanotubes in vacuo, as a function of separation $D$, both for strictly parallel
tubes of length 1 micron using Eq. (1) of \cite{Dobson06}, and for infinitely long 
tubes at angle $\theta = 1^0$ from Eq. (\ref{EMetFromh}).  
$F^{ins} \propto D^{-6}$ is the dispersion force from the remaining insulating  electron bands via the usual pairwise summation approach using data from
 \cite{Saito} and \cite{UnivGraphPot}. For comparison, $F^{e-e}$ gives the force between two localized (impurity) single-electron charges separated by $D$.  The dispersion force $F^{disp} = F^{met}+F^{ins}$ could be distinguished from that due to any localised extra electrons because $F^{disp}$, unlike $F^{e-e}$, will be invariant when one tube is made to slide along its own length.  
 \begin{figure}
\begin{tabular}{cccccc}
 \hfill \vline \,\vline & $F^{met}\hfill \vline $ & $F^{ins}\hfill \vline $ & $F^{met}\hfill \vline $ & $F^{ins}$ \hfill \vline & $F^{e-e}$ \\ 
  \hfill \vline \,\vline & L=1\,$\mu$  \hfill \vline & L=1\,$\mu$  \hfill \vline 
& $L\rightarrow \infty$ \hfill \vline  & $L \rightarrow \infty$ \hfill \vline& \\ 
D(nm)\hfill \vline \,\vline  & $\theta =0$ \hfill \vline & $\theta =0$ \hfill \vline & $\theta =1^{0}$ \hfill \vline & $\theta =1^{0}$  \hfill \vline & \\
 \hline \hline
2\, \hfill  \vline \,\vline & 250 \hfill \vline & 1370 \hfill \vline & 41\hfill \vline  & 205\hfill \vline  & 57.6 \\ 
5\,  \hfill \vline \,\vline & 9.3\hfill \vline  & 5.6 \hfill \vline  & 3.6 \hfill \vline &  2.1 \hfill \vline & 9.2 \\ 
10\, \hfill \vline \,\vline & 0.85 \hfill \vline  & 0.088 \hfill \vline  & 0.7  \hfill \vline  & 0.065\hfill \vline  & 2.3%
\end{tabular}
\caption{vdW Force contributions in picoNewtons between (5,5) CNTs (1st 4 columns) and Coulomb force between two electrons.}
\end{figure}
 

We note finally that the present theoretical results can be understood in terms of long-wavelength collective excitations, and are not limited to $T=0K$. However in practice $T$ needs to be low enough for a long electronic mean free path $> D$ to be maintained.

J. D. and I. K. acknowledge the hospitality of KITP
and partial support from the
NSF under Grant No. PHY05-51164. J. D. and T. G. were supported by the CSIRO National Hydrogen Materials Alliance.  Discussions with R. Podgornik, A. Parsegian, M. Kardar, T. Emig and R. French were much appreciated.


\end{document}